\documentclass[prl,bibnotes,twocolumn]{revtex4}
\begin{document}
\draft
\def\ds{\displaystyle}
\title{Berry's Phase for Ultracold Atoms in an Accelerated Optical Lattice}
\author{C. Yuce }
\address{ Physics Department, Anadolu University,
 Eskisehir, Turkey}
\email{cyuce@anadolu.edu.tr}
%\date{\today}
%\pacs{03.75.Lm, 03.65.Vf, 42.50.Wk, 37.10.Jk}
\begin{abstract}
Berry's phase is investigated for ultracold atoms in a frequency
modulated optical lattice. It is shown that Berry's phase appears
due to Bloch oscillation and the periodic motion of the optical
lattice. Particularly, Berry's phase for ultracold atoms under the
gravitational force in an oscillating tight-binding optical
lattice is calculated analytically. It is found that the Berry's
phase depends linearly on the amplitude of the oscillation of the
optical lattice.
\end{abstract}
\maketitle

\section{Introduction}

The dynamics of a charged particle in a one-dimensional spatially
periodic potential in the presence of a uniform electric field has
attracted considerable attention over the years. The dynamical
response of the particle to the electric field is the well-known
phenomena of Bloch oscillations. The researchers were unable to
observe Bloch oscillation in natural crystals since the scattering
time of the electrons by the lattice defects is much shorter than
the Bloch period. The fabrication of semiconductor superlattices
played an important role for the experimental verification of this
pure theoretical problem. The first experimental observation of
Bloch oscillation was reported for superlattices in 1992
\cite{bl0}. With the advent of laser cooling and trapping
techniques, it became possible to observe Bloch oscillation and to
check some other theoretical predictions experimentally in an
unprecedented way
\cite{deney1,deney2,deney3,deney4,deney5,bl2,bl}. For example,
Bloch oscillations was observed both with ultracold cesium atoms
\cite{deney1} and with a Bose-Einstein condensate \cite{deney2} in
an optical lattice. The experiment \cite{deney4} was the first one
with a Bose–Einstein condensate (BEC) of Rubidium atoms in an
optical lattice with gravity acting as the static field. The
persistent Bloch oscillations were observed for $\ds{t=10 s}$
using laser-cooled strontium atoms $\ds{^{88}Sr}$ in optical
lattices \cite{bl2}. Wannier-Stark ladders \cite{deney3} and
Landau-Zener tunneling were also observed experimentally \cite{deney5}. \\
In contrast to natural crystals, lattice constant, potential well
depth and lattice motion are controllable for optical lattices,
where standing laser waves and cold neutral atoms play the role of
the crystal lattices and the electrons, respectively. The motion
of the lattice is possible by generating frequency shift between
the two laser beams
\cite{bl,deney3,deney5,onemli,d00,d2,d3,d4,d5,d6,d7,d8,d9,d10,d11,olz1,olz2,olz3}.
If the phase difference is modulated periodically in time, the
wells in the optical lattice can be periodically shaken back and
forth. In \cite{deney3}, the motion of ultracold atoms in an
accelerating potential of the form $\ds{V_0 \cos (2 k_L~
(z-g(t))}$, where $\ds{g(t)=c_1t^2+c_2 \cos(\omega t)}$,
$\ds{c_1,c_2,\omega}$ are constants was considered. In the
reference frame of the standing wave, this potential becomes
$\ds{V_0 \cos (2 k_L~ z)+z+z\cos(\omega t)}$. In the rest frame of
the optical lattice, there will be an additional force acting on
the condensate atoms due to the oscillation of the lattice. The
rocking optical lattice was also used to demonstrate the dynamical
localization of matter-wave packets \cite{onemli}, which was
originally proposed by Dunlap and Kenkre for a charged particle in
a tight-binding lattice driven by a sinusoidal electric field
\cite{d0}. In the experiment \cite{olz1}, Wannier-Stark intraband
transitions was observed for ultracold atoms influenced by the
gravitational force
in a vertically oriented oscillating optical lattice. \\
In this paper, we will investigate one another physical phenomena
for an accelerated optical lattice. This is the Berry's phase
\cite{berry,berry2,berryrev}. The concept of the Berry's phase is
of great interest in a variety of different branches of physics.
Of particular example in solid state physics is the Berry's phase
for the problem of motion of an electron in a periodic potential
and a time-dependent electric field \cite{zak,zak2,zakgibi}. Here
we will derive a formula for the Berry's phase for neutral atoms
placed in a periodically shaken
optical lattice.\\
The paper is structured as follows. The following section briefly
reviews the concept of the Berry's phase. The final section
studies the Berry's phase for ultracold atoms in an accelerated
optical lattice.

\section{Berry's Phases }

In 1984, Berry reinvestigated an old problem of adiabatic
evolution of a quantum state when the time dependent external
parameters change periodically \cite{berry}. He considered a time
periodic Hamiltonian and supposed that the external parameters are
slow enough so that there is no transition to the higher energy
levels. He found that in the absence of degeneracy, the eigenstate
for a cyclic Hamiltonian comes back to itself after a period but
takes an extra phase difference. This phase difference, later
commonly called Berry's phase, is gauge invariant. The gauge
invariance property makes the Berry's phase physical. The Berry's
phase is also geometrical, that is, it doesn't depend on the exact
rate of change of the external parameters in the Hamiltonian. So,
the Berry's phase
is expressed in terms of local geometrical quantities. \\
Let us introduce the basic concepts of the Berry phase arising
from the adiabatic evolution of a quantum state. Consider a
physical system described by a Hamiltonian that depends on time.
Suppose that the Hamiltonian is cyclic with a period $\ds{T}$.
\begin{equation}\label{fhdy}
H(t+T)=H(t)~.
\end{equation}
We are interested in the adiabatic evolution of the system. Hence,
the system initially in one of the eigenstates of the Hamiltonian
$\ds{H(0)}$ will stay as an instantaneous eigenstate of the
Hamiltonian throughout the process. The only difference is the
phase difference between the initial and final quantum states. The
wave function for a time periodic Hamiltonian (\ref{fhdy}) at time
$\ds{t}$ reads
\begin{equation}\label{tf0976td}
\Psi(z,t)=\exp{\left(i\gamma(t)-\frac{i}{\hbar}\int{Ed{t}}\right)}~\phi(z,t)~,
\end{equation}
where $\ds{E}$ is the corresponding time dependent energy, the
function $\ds{\phi(z,t)}$ satisfies
\begin{equation}\label{ghdf56s}
H\phi(z,t)=E\phi(z,t)~,
\end{equation}
and $\ds{\gamma(t)}$is the Berry's phase, which was usually
neglected in the theoretical treatment of time-dependent problem
until Berry reconsidered the cyclic evolution of the system.
Substitution of (\ref{tf0976td},\ref{ghdf56s}) into the
corresponding Schrodinger equation yields
\begin{equation}\label{gamma}
\dot{\gamma}=i\hbar<\phi(z,t)|\frac{\partial{\phi(z,t)}}{\partial
{t}}>~,
\end{equation}
where dot denotes time derivation. It was generally assumed that
$\ds{\gamma(t)}$ could be eliminated by redefining the phase of
the eigensate. Berry, however, realized that such a phase is
observable when the system comes to its initial state.
\begin{equation}\label{gamma2}
\gamma=i\hbar\oint{d{t}}<\phi(z,t)|\frac{\partial{\phi(z,t)}}{\partial
{t}}>~.
\end{equation}
Having briefly reviewed the Berry's phase, let us study the
Berry's phase for ulrtacold atoms in an accelerated optical
lattice.

\section{Formalism}

Consider an atom influenced by a time dependent linear potential
in an optical lattice. Suppose that the optical lattice is shaken
by a frequency shift between the two lattice beams. We will
investigate the Berry's phase for such a system. The optical
lattice is no longer stationary in space. We assume that the
transverse motion of the atoms is frozen (i.e., we are dealing
with a 1-D problem). Then the corresponding Hamiltonian reads
\cite{olz1,olz2,olz3}
\begin{equation}\label{hamilt}
H=\frac{p^2}{2m} +V_0 \cos \left\{2 k_L~
(z-z_0(t))\right\}+f(t)~z~,
\end{equation}
where $\ds{m}$ is the atomic mass, $\ds{V_0}$ is the lattice
depth, $\ds{k_L}$ is the optical lattice wave number, $\ds{f(t)}$
is, in general, a time-dependent force and $\ds{z_0(t)}$ is a
time-dependent periodic phase.\\
We are interested in the adiabatic evolution of the system. In
other words, the rate of change of $\ds{z_0(t)}$ and $\ds{f(t)}$
are much slower than frequency between the bands so that it
doesn't induce interband transition. From the experimental point
of view, the requirement for $\ds{z_0(t)}$ is practically met by
having a much smaller frequency shift between the lattice beams
than the relevant band gaps. The requirement for $\ds{f(t)}$ can
also be experimentally reached. As a particular example, the
gravitational force acting on neutral atoms is a good choice to
observe Berry's phase experimentally.\\
We further assume that $\ds{f(t+T)=f(t)}$ and
$\ds{z_0(t+T)=z_0(t)}$, where $\ds{T}$ is the period so that the
Hamiltonian is time periodic. We emphasize that there exists one
another parameter that must be periodic with the same period
$\ds{T}$. It is the quasimomentum, $\ds{k}$. The periodicity of
$\ds{k}$ with the period $\ds{T}$ is required since the path must
be closed to make Berrys' phase a gauge-invariant quantity with
physical significance. The linear potential in (\ref{hamilt})
makes the quasimomentum $\ds{k}$ vary over the entire Brillouin
zone and generates a closed path in the momentum space. Here we
are mainly interested in the dynamics of Bloch oscillations and
therefore use a weak force and initial states populating the
lowest Bloch band. Unless the acceleration and the force is large
enough for the atoms to undergo a Landau-Zener tunneling, atoms
will remain in the first band.\\
The time-dependent Schrodinger equation corresponding to the
Hamiltonian (\ref{hamilt}) is given by
\begin{equation}\label{tjhjt}
i\hbar \frac{\partial \Psi}{\partial t}=\left(-\frac{\hbar^2}{2m}
\frac{\partial^2 }{\partial z^2}+V_0 \cos \left\{2 k_L~
(z-z_0(t))\right\} +f(t) z\right)\Psi~.
\end{equation}
In the adiabatic limit, the wave function $\ds{\Psi(z,t)}$ is
given by the equation (\ref{tf0976td}). Now, we need
$\ds{\phi(z,t)}$, which satisfies the equation (\ref{ghdf56s})
with the Hamiltonian (\ref{hamilt}). The energy spectrum of an
atom in a periodic potential consists of the Bloch bands.
According to Bloch's theorem, the eigenstates of a periodic
Hamiltonian is given by
\begin{equation}\label{bloch}
\phi(z,t)=e^{ikz } u_{n,k} (z,t)~.
\end{equation}
where $\ds{k}$ is the quasimomentum and $\ds{u_{n,k} (z,t)}$
satisfies the following equation
\begin{equation}\label{cem01}
\left(\frac{(p+{\hbar}k)^2}{2m}+V_0 \cos \left\{2 k_L~
(z-z_0(t))\right\}+f(t) z\right)u=Eu~.
\end{equation}
Note that the energy $\ds{E}$ is, in general, time dependent. From now on, we will drop the index $\ds{n}$.\\
The Berry's phase for our system becomes
\begin{equation}\label{gamma2}
\gamma=i{\hbar}\oint{d{t}}<\phi|\frac{\partial{\phi}}{\partial
{t}}>=i{\hbar}\int_0^T{d{t}}<\phi|\frac{\partial{\phi}}{\partial
{t}}>~.
\end{equation}
Note that the Berry's phase is independent of how the time
dependent periodic functions, $\ds{z_0(t)}$ and
$\ds{f(t)}$, vary in time.\\
Since the system is oscillating in time, one should be careful
when calculating the scalar product
$\ds{<\phi|\frac{\partial{\phi}}{\partial {t}}>}$ in the above
integral. It is preferable to transform the moving frame to the
stationary lattice frame. The coordinate transformation is
\begin{equation}\label{s9dsf}
z^{\prime}=z-z_0(t)~.
\end{equation}
Let us rewrite the equation (\ref{cem01}) in the lattice
coordinate frame. It becomes
\begin{equation}\label{cem074e}
\left(\frac{(p+{\hbar} k)^2}{2m}+V_0 \cos \left(2 k_L
z^{\prime}\right) +f(t) z^{\prime}
\right)u^{\prime}=E^{\prime}~u^{\prime}~.
\end{equation}
where $\ds{E^{\prime}=E-z_0(t)f(t)}$ and $\ds{u^{\prime}=u(z-z_0(t))}$.\\
In the absence of the force, $\ds{f(t)\rightarrow0}$, $\ds{k}$ is
a conserved quantity and the Bloch function is specified by a band
index $\ds{n}$ and $\ds{k}$. In the presence of the additional
force $\ds{f(t)}$, the wave packet can be characterized by a
single mean quasimomentum $\ds{k(t)}$ at time $\ds{t}$ if the
width of the wave packet in quasimomentum space is small. The wave
packet prepared with a well-defined quasimomentum will oscillate
in position. Applying a perturbation to the atom make $\ds{k}$
vary on a closed path in the Brillouin zone according to the
equation
\begin{equation}\label{bloch2}
\dot{k}=\frac{f(t)}{\hbar}~.
\end{equation}
In a one dimensional system, $\ds{k}$ sweeps the interval
$\ds{(-\pi/a,\pi/a)}$, where $\ds{a}$ is the lattice constant.\\
Under the coordinate transformation (\ref{s9dsf}), Berry's phase
is also transformed. Note that the time derivative operator
transforms as $\ds{\frac{\partial}{\partial
{t}}\rightarrow\frac{\partial}{\partial
{t}}-\dot{z_0}\frac{\partial}{\partial {z^{\prime}}}}$. Hence the
Berry's phase (\ref{gamma2}) is transformed according to
\begin{equation}\label{gamma3ek}
\gamma=i\hbar\oint{d{t}}<\phi^{\prime}|\frac{\partial{\phi^{\prime}}}{\partial
{t}}>-i\hbar\oint{d{t}}~\dot{z_0}~<\phi^{\prime}|\frac{\partial{\phi^{\prime}}}{\partial
{z^{\prime}}}>~,
\end{equation}
where $\ds{\phi^{\prime}=e^{ikz } u^{\prime}}$. Using
(\ref{bloch2}) and $\ds{\oint{f(t)~d{t}} =0}$, we get
\begin{equation}\label{gamma3}
\gamma=i\hbar\oint{d{t}}<u^{\prime}|\frac{\partial{u^{\prime}}}{\partial
{t}}>-i\hbar\oint{d{t}}~\dot{z_0}~<\phi^{\prime}|\frac{\partial{\phi^{\prime}}}{\partial
{z^{\prime}}}>~,
\end{equation}
The scalar products in the above integral will be calculated in
the stationary lattice frame. Let us study the Berry's phase
(\ref{gamma3}) in detail. There are two terms in the Berry's
phase. The first term appears because the quasimomentum $\ds{k}$
is made to vary across the Brillouin zone. The first term exists
even if the optical lattice is stationary. It was studied before
and known as Zak's phase \cite{zak}. The value of the Zak's phase
depends upon the symmetry of the lattice. It is usually found to
be either zero or $\ds{\pi}$ in the presence of inversion
symmetry. Defining
\begin{equation}\label{g55}
q_n=\frac{a}{2\pi}\int_{-\pi/a}^{\pi/a}\chi_{nn}(k)~dk~.
\end{equation}
where
\begin{equation}\label{gfsdy4}
\chi_{nn}(k)=i\hbar\frac{2\pi}{a} \int_0^a u_{nk}^{\star}
\frac{{\partial}u_{nk}}{{\partial}k}~dx.
\end{equation}
the Zak's phase is given by  \cite{zak}
\begin{equation}\label{dfh54r5}
\gamma_{Zak}=i\hbar\oint{d{t}}<\phi^{\prime}|\frac{\partial{\phi^{\prime}}}{\partial
{t}}>=\frac{2\pi}{a} q_n~.
\end{equation}
The second term in (\ref{gamma3}) is due to the fact that the
optical lattice is oscillating in time. It can be further
simplified if we use the relation;
$\ds{-i{\hbar}<\phi^{\prime}|\frac{\partial{\phi^{\prime}}}{\partial
{z^{\prime}}}>=p^{\prime}_{A}}$, where $\ds{p^{\prime}_{A}}$ is
the wavepacket momentum for atoms in the stationary frame;
$\ds{p^{\prime}_{A}=\frac{\partial E^{\prime}}{\partial{k}}}$. The
rate of change of the quasi momentum is given by the external
force (\ref{bloch2}), while the rate of change of the momentum is
given by the total force including the influence of the periodic
potential. Using $\ds{p^{\prime}_{A}=mv^{\prime}_{A}}$ and
$\ds{v_{L}=\dot{z_0}}$, where $\ds{v_{L}}$ is the velocity of the
optical lattice, the equation (\ref{gamma3}) is reduced to
\begin{equation}\label{gammason}
\gamma=\gamma_{Zak}+m\oint{d{t}}~v_{L}~ v^{\prime}_{A}~.
\end{equation}
It is the Berry's phase for the Hamiltonian (\ref{hamilt}). The
additional Berry's phase due to the oscillation of the optical
lattice is equal to the time integral of the momentum of atoms
times the velocity of the optical lattice. This second term is, in
general, nonzero as can be seen for a specific example given below.\\
As an example, let us suppose that the force is equal to a
constant, $\ds{f(t)=f_0=mg}$, where $\ds{m}$ and $\ds{g}$ are mass
and the gravitational acceleration, respectively. Such a constant
force always exists in the optical lattice because the atoms feels
the gravitational force. The solution of (\ref{bloch2}) gives the
time evolution of the quasimomentum. It is given by
\begin{equation}\label{period0}
k(t)=k_0+\frac{f_0}{\hbar}t~,
\end{equation}
where $\ds{k_0}$ is the initial value of the quasimomentum. The
quasi momentum varies linearly with time for the constant force.
The period of the motion in $\ds{k}$-space is given by
$\ds{T=\frac{2 \pi
\hbar}{af_0}}$, where $\ds{a}$ is the lattice constant. \\
Suppose now that the optical lattice is periodically kicked, that
is
\begin{equation}\label{gt}
z_0(t)=L~\sin{\omega t}~.
\end{equation}
where the amplitude $\ds{L}$ is a constant and the angular
frequency $\ds{\omega=af_0/\hbar}$ ( the period of the oscillation
is equal to  the period of the motion in $\ds{k}$-space). The
system comes to its initial position after a period $\ds{T}$. The
velocity of the optical lattice is $\ds{v_{L}= {\omega}{L}\cos{\omega t}}$.\\
We need the momentum, $\ds{p^{\prime}_A}$, to calculate the
Berry's phase analytically (\ref{gammason}). Let us use the tight
binding approximation. In this case, the energy is given by
$\ds{E^{\prime}=E_0-2 \delta \cos{k a}}$, where $\ds{E_0}$ and
$\ds{\delta}$ are constants. Under the action of the force, the
particle state changes continually, so does the momentum
$\ds{p^{\prime}_A=-2 a \delta \sin{k a}}$. If we use the relation
(\ref{period0}), we get $\ds{p^{\prime}_A=-2 a \delta \sin{(k_0
a+{\omega}t) }}$. Substitution of $\ds{v_{L}}$ and
$\ds{p^{\prime}_{A}}$ expressions into the equation
(\ref{gammason}) yields
\begin{equation}\label{cemyuce0}
\gamma=\gamma_{Zak}+\gamma_{ocs.}
\end{equation}
where $\ds{\gamma_{ocs.}}$ is the Berry's phase due to the
oscillation of the optical lattice
\begin{equation}\label{cemyuce}
\gamma_{osc}=-2{\pi}m {\delta}  aL \sin(k_0 a)
\end{equation}
This is the Berry's phase for the constant force in a periodically
kicked tight-binding optical lattice. \\
When there is no oscillation, $\ds{L{\rightarrow}~0}$, the Berry's
phase is just equal to the Zak's phase. In the rocking optical
lattice, the Bloch state picks up an additional phase
(\ref{cemyuce}). The Berry's phase, which increases linearly with
the amplitude $\ds{L}$, can be controlled by changing the
amplitude of the oscillation. However, it does not depend on how
$\ds{z_0(t)}$ changes in time. It depends on the geometry: the
lattice length $\ds{a}$, the amplitude $\ds{L}$ and the initial
value of the
quasimomentum $\ds{k_0}$.\\
We would like to point out that different energy bands have
different Berry's phases. In principle, any interference
experiment with neutral atoms for two
such energy bands enables one to measure the phase difference.\\
One can eliminate the Zak's phase by using two independent
Bose-Einstein condensates of the same atomic species and the same
band indexes \cite{interfer}. In this case, Zak's phases for the
two independent condensates are the same, while the Berry's phases
due to the oscillation, $\ds{\gamma_{osc}}$, are, in general,
different. Particularly, if the two independent condensates are
oscillated with different amplitudes, the phase difference due to
the oscillation of the lattice can be measured. So, the
theoretical results for $\ds{\gamma_{osc}}$ can be checked by such
an interference experiment. Particularly, $\ds{^{88}Sr}$ atoms is
a good choice for such an experiment because it has remarkably
small atom-atom interactions and in the ground state it has zero
orbital, spin and nuclear angular momentum.\\
We would like to prove that the Berry's phase (\ref{cemyuce0}) is
gauge invariant. By a gauge transformation, $\ds{\Psi=e^{-iA
z/{\hbar}}~ \psi}$, where $\ds{A=f_0 t}$ is the gauge potential,
the equation (\ref{tjhjt}) is transformed to
\begin{equation}\label{tjhjt02}
i\hbar \frac{\partial \psi}{\partial t}=\left(\frac{(p-A)^2}{2m}
+V_0 \cos \left\{2 k_L~ (z-z_0(t))\right\} \right)\psi~.
\end{equation}
Therefore the Berry's phase (\ref{gamma2}) is transformed
according to
\begin{equation}\label{gammagauge}
\gamma^{\prime}=\gamma+f_0\oint{d{t}}<z>
\end{equation}
where $\ds{\gamma^{\prime}}$ is the transformed Berry's phase
under the gauge transformation and $\ds{<z>=<\phi|z|\phi>}$. Using
the relation $\ds{<z>=<z^{\prime}>+z_0}$, where $\ds{
\frac{d<z^{\prime}>}{dt}=mv^{\prime}_A }$ and the expression for
$\ds{z_0(t)}$ (\ref{gt}), one can readily prove that the second
term in the right hand side of (\ref{gammagauge}) vanishes. Hence,
\begin{equation}\label{gammagauge2}
\gamma^{\prime}=\gamma~.
\end{equation}
The gauge invariance of the Berry's phase guarantees that it can
be observable.\\
In conclusion, we have shown that there exists non-vanishing
Berry's phase for the rocking optical lattice with the linear
potential. The Berry's phase consists of the two terms,
$\ds{\gamma_{Zak}}$ and $\ds{\gamma_{osc}}$. The first one was
already studied by Zak. In this case, the parameter space (the
Brillouin zone) exists naturally \cite{zak}. The second one,
$\ds{\gamma_{osc}}$, arises due to the oscillation of the optical
lattice. As an example, the Berry's phase for ultracold atoms in a
periodically kicked tight-binding optical lattice is calculated
analytically. It is found that the Berry's phase
$\ds{\gamma_{osc}}$ can be controlled by the amplitude of the
oscillation. However, the frequency of the oscillation has no
effect on it. We have also discussed that the Berry's phase can be
measured with the current technology by an interference experiment
with two independent optical lattices. Here, our investigation is
restricted to the one dimension. This is not only for
simplification. The first example that Berry's phase appears in
one dimension was given by Zak. Here, another example in one
dimension has been given. The generalization of the results to
three dimensional systems is straightforward. Finally, we have
proven that the Berry's phase remains invariant under a gauge
transformation.\\
We would like to acknowledge useful discussions with Yuksel
Ergun.\\


\begin{thebibliography}{99}
\bibitem{bl0} J. Feldmann, K. Leo, J. Shah, B. A. B. Miller, J. E. Cunningham, T. Meier, G. von Plessen, A. Schulze, P. Thomas, and S. Schmitt-Rink, Phys. Rev. B {\bf 46}, 7252 (1992).
\bibitem{deney1} M. Ben Dahan, E. Peik, J. Reichel, Y. Castin, and C. Salomon, Phys. Rev. Lett. {\bf 76}, 4508 (1996).
\bibitem{deney2} O. Morsch, J. H. Muller, M. Cristiani, D. Ciampini, and E. Arimondo, Phys. Rev. Lett. {\bf 87}, 140402 (2001).
\bibitem{deney4} B. P. Anderson and M. Kasevich, Science {\bf 282}, 1686 (1998).
\bibitem{bl2} G. Ferrari, N. Poli, F. Sorrentino, and G. M. Tino, Phys. Rev. Lett. 97,  {\bf 060402} (2006).
\bibitem{deney3} S. R. Wilkinson, C.F. Bharucha, K.W. Madison, Q. Niu, and M. G. Raizen, Phys. Rev. Lett. {\bf 76}, 4512 (1996).
\bibitem{deney5} A. Zenesini, H. Lignier, G. Tayebirad, J. Radogostowicz, D. Ciampini, R. Mannella, S. Wimberger, O. Morsch, E. Arimondo, Phys. Rev. Lett.  {\bf 103}, 090403 (2009).
\bibitem{bl} O. Morsch, J. H. Müller, M. Cristiani, D. Ciampini, and E. Arimondo, Phys. Rev. Lett.  {\bf 87}, 140402 (2001).
\bibitem{onemli} A. Eckardt, M. Holthaus, H. Lignier, A. Zenesini, D. Ciampini, O. Morsch, and E. Arimondo, Phys. Rev. A {\bf 79}, 013611 (2009).
\bibitem{d00} K. W. Madison, M. C. Fischer, and M. G. Raizen, Phys. Rev. A {\bf60}, R1767 (1999).
\bibitem{d2} Q. Thommen, J. C. Garreau,  Z. Veroniquei, Phys. Rev. A {\bf 65}, 053406 (2002).
\bibitem{d3} Sierk Pötting, Marcus Cramer, Christian H. Schwalb, Han Pu, and Pierre Meystre, Phys. Rev. A {\bf64}, 023604 (2001).
\bibitem{d4}K. W. Madison, M. C. Fischer, R. B. Diener, Qian Niu, and M. G. Raizen, Phys. Rev. Lett. {\bf 81}, 5093 (1998).
\bibitem{d5} N. Gemelke, E. Sarajlic, Y. Bidel, S. Hong, and S. Chu, Phys. Rev. Lett. {\bf 95}, 170404 (2005).
\bibitem{d6} Wenhua Hai, Gengbiao Lu, and Honghua Zhong, Phys. Rev. A  {\bf 79}, 053610 (2009).
\bibitem{d7} F. Saif and I. Rehman, Phys. Rev. A  {\bf 75}, 043610 (2007).
\bibitem{d8} M. C. Fischer, K. W. Madison, Qian Niu, and M. G. Raizen, Phys. Rev. A  {\bf 58}, R2648 (1998).
\bibitem{d9} M. Cristiani, O. Morsch, J. H. Müller, D. Ciampini, and E. Arimondo, Phys. Rev. A {\bf65}, 063612 (2002).
\bibitem{d10} Yi Zheng, Marijan Kostrun, and Juha Javanainen, Phys. Rev. Lett. {\bf 93}, 230401 (2004).
\bibitem{d11} Thawatchai Mayteevarunyoo and Boris A. Malomed, Phys. Rev. A {\bf  80}, 013827 (2009).
\bibitem{olz1} V. V. Ivanov, A. Alberti, M. Schioppo, G. Ferrari, M. Artoni, M. L. Chiofalo, and G. M. Tino, Phys. Rev. Lett. {\bf 100}, 043602 (2008).
\bibitem{olz2} Q. Thommen, J. C. Garreau, and V. Zehnle, J. Opt. B: Quantum Semiclass. Opt.  {\bf 6} 301 (2004).
\bibitem{olz3} Alessandro Zenesini, Hans Lignier, Donatella Ciampini, Oliver Morsch, and Ennio Arimondo, Phys. Rev. Lett. {\bf 102}, 100403 (2009).
\bibitem{d0} D. H. Dunlap and V. M. Kenkre, Phys. Rev. B  {\bf 34}, 3625 (1986).
\bibitem{berry} M. V. Berry, Proc. R. Soc. London Ser. A {\bf 392}, 45 (1984).
\bibitem{berry2} Geometric Phases in Physics, edited by A. Shapere, F. Wilczek (World Scientific, Singapore, 1989).
\bibitem{berryrev} Di Xiao, Ming-Che Chang, Qian Niu, arXiv:0907.2021.
\bibitem{zak} J. Zak, Phys. Rev. Lett. {\bf 62}, 2747 (1989).
\bibitem{zak2} L. Michel, J. Zak, Europhys. Lett. {\bf 18}, 239 (1992).
\bibitem{zakgibi} M. J. Rave and W. C. Kerr, Eur. Phys. J. B {\bf 45}, 473 (2005).
\bibitem{interfer} M. Naraschewski et al., Phys. Rev. A {\bf 54}, 2185 (1996).
\end{thebibliography}
\end{document}